\documentclass[]{tMOP2e}

\usepackage{subfigure}

\theoremstyle{plain}

\theoremstyle{remark}

\begin{document}



\title{Nanoantennas for enhanced light trapping in transparent organic solar cells}

\author{P.M. Voroshilov$^{\rm a}$$^{\ast}$\thanks{$^\ast$Corresponding author. Email: pavel.voroshilov@phoi.ifmo.ru\vspace{6pt}}, C.R. Simovski$^{\rm a,b}$,  and P.A. Belov$^{\rm a}$\\\vspace{6pt}
$^{a}${\em{ITMO University, St. Petersburg 197101, Russia}};\\
$^{b}${\em{Aalto University, School of Electrical Engineering, Department of Radio Science and Engineering, P.O. Box 13000, 00076 Aalto, Finland}}
}

\maketitle

\begin{abstract}
We propose a light-trapping structure offering a significant enhancement of photovoltaic absorption in transparent organic solar cells operating at infrared while the visible light transmission keeps sufficiently high. The main mechanism of light trapping is related with the excitation of collective oscillations of the metal nanoantenna arrays, characterized by advantageous field distribution in the volume of the solar cell. It allows more than triple increase of infrared photovoltaic absorption.
\end{abstract}


\section{Introduction}

In recent decades, a rapid growth occurs in one of the most promising alternative energy industry -- photovoltaics (PV). Organic photovoltaic (OPV) cells based on conducting polymers and organic molecules belong to the so-called third generation of solar cells. Combined with high sensitivity to the non-direct (e.g. scattered by clouds or inhomogeneities of the atmosphere) and low-intensity solar irradiance, flexibility and light weight, panels of OPV cells have low production cost especially in comparison with traditional solar cells. These advantages potentially make them economically viable for many applications \cite{Luke}.

Transparency in the wavelength range of visible light is one of the interesting properties of certain OPV materials. This feature extends the applications of OPV cells beyond those pointed out in \cite{Luke}. For example, solar cells operating in the infrared or ultraviolet ranges can be integrated into windows of residential and office buildings, automobile glasses, touchscreens of smart devices etc.  The visible solar light transmits inside and the infrared radiation is converted to electricity. Notice, that the total transmission of visible light is practically not required, one may sacrifice certain decrease of transmittance in favor of the electric energy generation. Really, a typical residential architectural window transmits nearly 55\% of the visible light energy \cite{glass}. 30\% of the visible solar light transmitted through the window is often sufficient for practical needs. In other words the solar cell located on the window or integrated in it should transmit more than one half of the incident visible light. Currently, there are several designs of transparent solar cells which transmits more than one half of incident solar radiation in the visible range. These include organic solar cells based on organic polymers \cite{polymer, polymer2} or molecular heterojunctions \cite{bulovic}, as well as inorganic heterojunction solar cell based on p-NiO and n-ZnO \cite{ultraviolet}.

OPV cells based on molecular heterojunctions may operate in the near infrared (NIR) range where the solar power spectrum is much higher than in the ultraviolet one and therefore promise quite high electric output. However, these cells have serious limitations on the thickness of their active layer. Due to a short exciton diffusion length (approximately 10-15 nm) in organic materials \cite{exciton}, thickness of active layer typically should be the same order, so that excitons could reach the heterojunction interface and break up into free carriers. However, such value of thickness is far from being sufficient for efficient absorption of infrared light. Therefore the overall efficiency in the operational band is very insufficient and the power output very low. Multilayer antireflection coatings (ARC) and distributed Bragg reflectors (DBR) combined with the bottom electrode were devloped for such solar cells. The photoelectric conversion efficiency of a transparent OPV cell can be improved using such multilayer structures by 30\% in the best cases \cite{bulovic}. This solution is quite expensive since requires the sub-nm precision for every nanolayer, and the modest enhancement of the efficiency hardly justifies the increase of fabrication costs.

In this paper, we consider the possibility of light trapping by arrays of metal nanoantennas \cite{nanoantennas} located on a transparent organic solar cell based on molecular heterojunction \cite{bulovic}, where the main absorption occurs in the NIR range. The mechanism of the light trapping is related with the excitation of so-called \textit{domino modes} in silver nanoantenna arrays, which represent non-plasmonic collective oscillations in array of metal nanostrips or nanobars. In the frequency range of domino modes the solar energy is transformed into a set of hot spots partially located ins between the metal elements, partially in their substrate with negligible penetration of enhanced electric field inside the metal. The penetration depth of the local field into metal is much smaller than the skin-depth and the incident light energy is practically not dissipated in the metal. The array supporting these modes operates beyond plasmon resonances as if metal elements were perfectly conducting metal. These modes are observed only for substantial metal nanobars or nanostrips whose thickness exceeds the skin-depth. A part of every hot spot located inside theactive layer corresponds to the enhanced power absorption which decreases the transmission of light in the operation band. In the spectrum of enhanced absorption the reflectance is also suppressed.  

\section{Structure and methods}

In our simulations, we considered the design of an OPV cell first suggested in \cite{bulovic} which is nearly transparent for the visible light. Molecular heterojunction is here formed by molecular organic donors, namely chloroaluminum phthalocyanine (ClAlPc), and molecular acceptor fullerenes $C_{60}$. A modification of this design was suggested in \cite{patent}: tin-phthalocyanine (SnPc) can serve as a donor. We have chosen this design mainly because we have not found in the available literature optical constants for ClAlPc, whereas those of SnPc are known for both NIR and visible ranges.  Also, SnPc has a maximum of absorption in the near infrared at wavelengths $\lambda$ = 700-950 nm and very low attenuation at wavelengths $\lambda$ = 420-600 nm. It provides a good transparency of the solar cell in the visible range. Solar cells based on SnPc (enhanced by ARC and DBR) show an overall efficiency nearly equal to 2\% and an averaged visible transmittance (AVT) that exceeds 70\% \cite{patent}.

\begin{figure}[h]
\begin{center}
\begin{minipage}{130mm}
\subfigure[Cross-section view of the OPV cell.]{
\resizebox*{4.5cm}{!}{\includegraphics{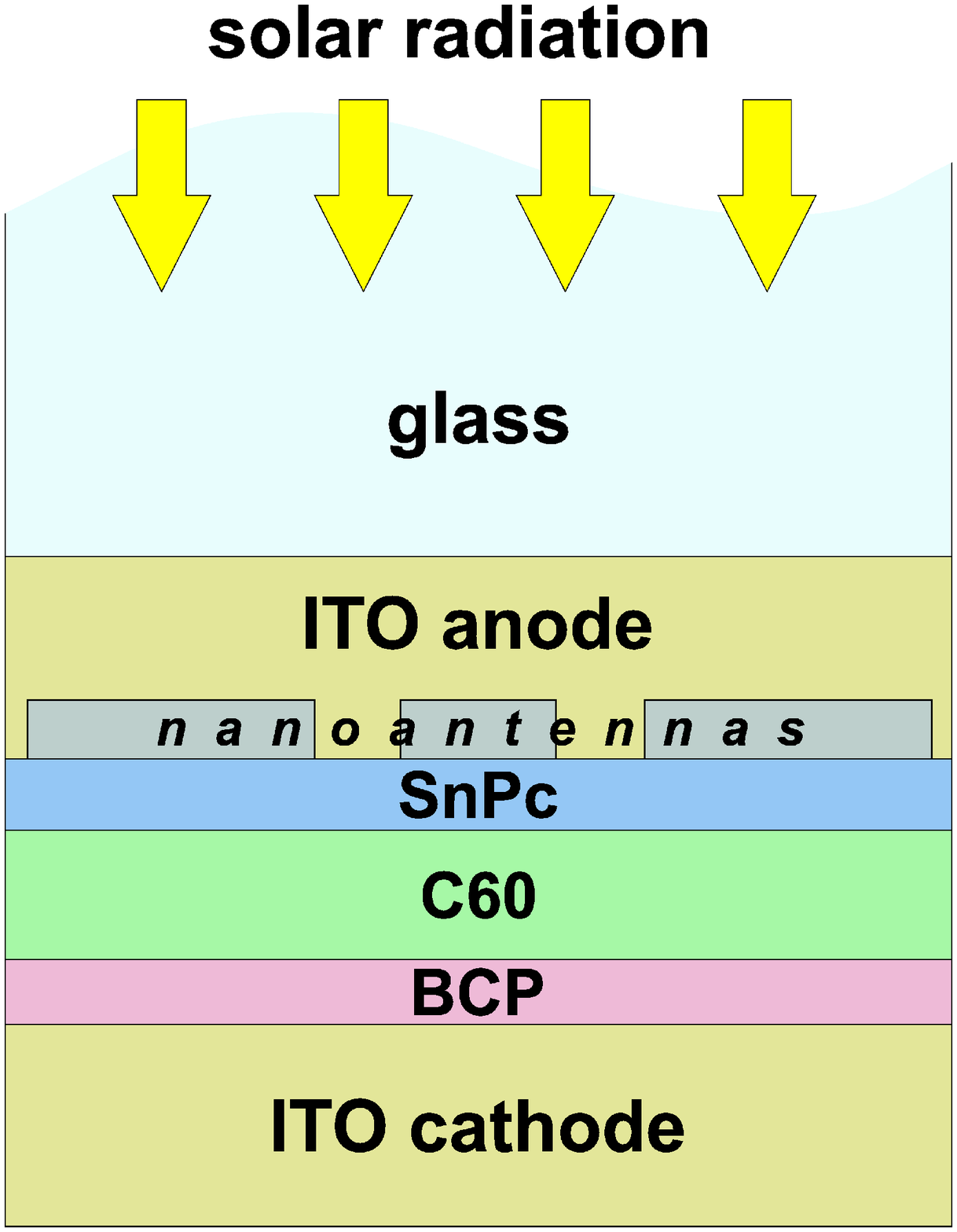}}}\hspace{6pt}
\subfigure[Nanoantennas arrangement in the unit cell.]{
\resizebox*{7.5cm}{!}{\includegraphics{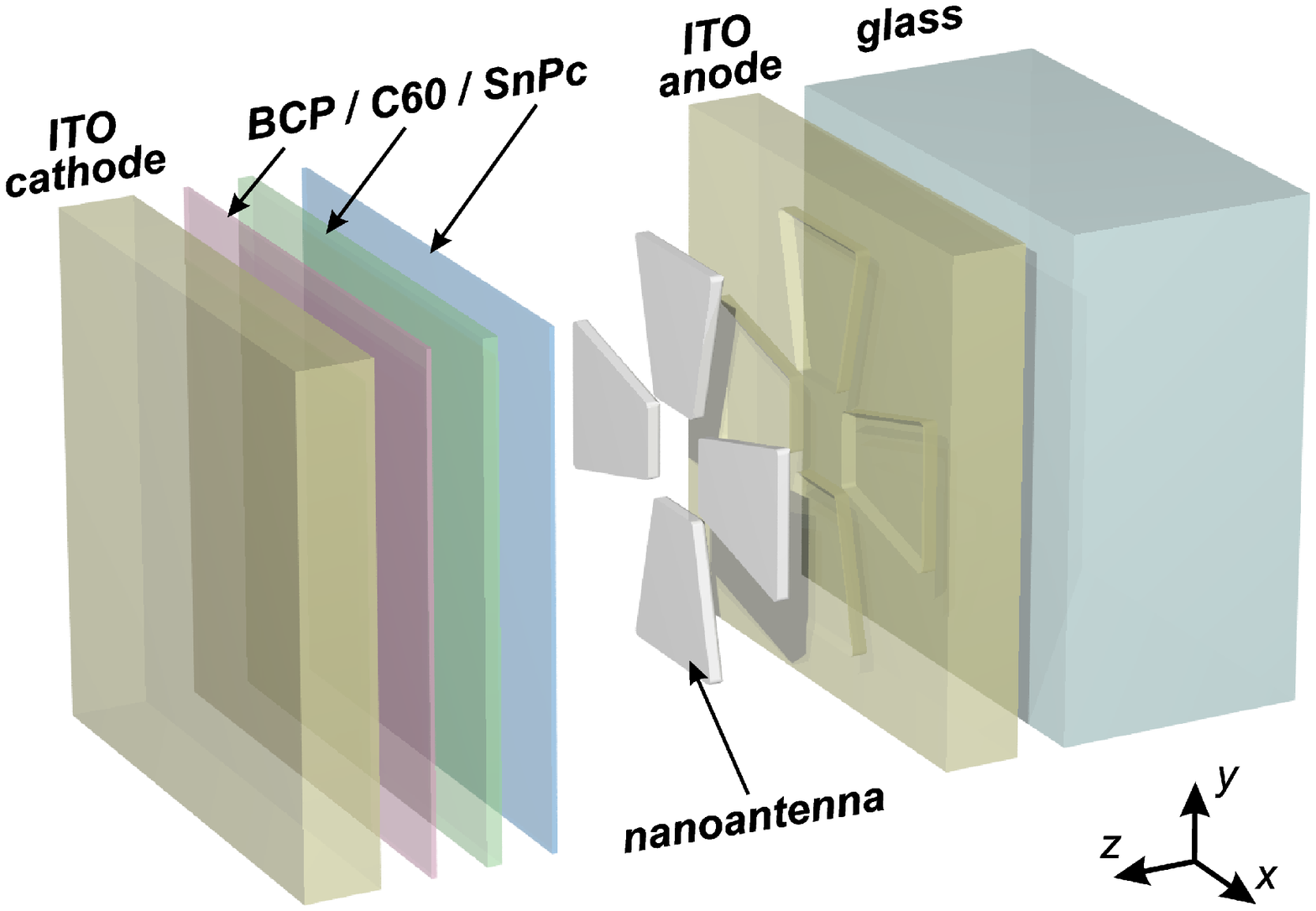}}}
\caption{\label{fig1} Schematic views of the transparent organic solar cell proposed by \cite{patent} with our light trapping structure (LTS). The solar cell created on a glass substrate is composed of the following layers: ITO (150 nm)/SnPc(10 nm)/$C_{60}$(30 nm)/BCP(10 nm)/ITO(initially 10 nm). Metal nanoantennas are embedded in the ITO anode. (The color version of this figure is included in the online version of the journal.)}
\label{pic1}
\end{minipage}
\end{center}
\end{figure}

The photovoltaic cell in our simulations is located on a glass substrate coated with a layer of indium tin oxide (ITO) which operates as a transparent anode. A layer of organic semiconductor bathocuproine (BCP) is placed between an acceptor $C_{60}$ and transparent ITO cathode layers. BCP facilitates the transfer of electrons to the cathode \cite{bulovic,exciton}. Nanoantennas realized in the form of tapered nanopatches which support domino modes in a rather broad frequency range were suggested in \cite{nanoantennas}. However, in that work nanoantennas of this geometry were used on a more substantial (100-150 nm) substrate of doped semiconductor. Their operation in the present case of a 40 nm thick SnPc/C$_{60}$ layer represents a new problem. The excitation of domino-modes on a so tiny photovoltaic substrate is not evident, and if they are excited their frequency range needs to be studied. Another novelty of the present work compared to \cite{nanoantennas} is our suggestion to incorporate nanoantennas into the anode of the solar cell. Silver elements are assumed to be covered with a 2-3 nm thick deposited layer of SiO$_2$. A so thin dielectric is sufficient to prevent the ohmic contact of nanoantennas with ITO but does not affect the electromagnetic fields and can be neglected in full-wave simulations.

A unit volume of the OPV device split in accord to the dimensions of the unit cell of our nanoantenna light trapping structure (LTS) is shown in Fig. \ref{pic1}. It is clear that we imply our solar cell located on the internal side of the window. Numerical simulations for a solar cell with LTS were carried out using Frequency Domain Solver in commercial package CST Microwave Studio. Periodic boundaries with two symmetry planes were set in \textit{x} and \textit{y} directions, and open boundary conditions (PML) were put in top and bottom along the z-axis. In our calculations a solar cell was illuminated by a normally incident wave with uniform power spectrum in the range $\lambda=$400-1000 nm. We calculated the useful absorption coefficient -- the power absorbed (per unit area) in the photovoltaic layer over the whole operation band of the solar cell (600-1000 nm) normalized to the power of the incident light. We also calculated the transmission coefficient in the range of the visible light (400-750 nm). To check the excitation of domino-modes and to outline their range we also calculated the spatial distributions of the electric field amplitude. 
The transfer matrix method was used to exactly calculate absorbtion and transmission spectra in the case of the pure solar cell -- that without LTS. Complex refractive indices of BCP and of glass, as well as optical constants of ITO were taken from \cite{materials}. Complex refractive index of SnPc was taken from \cite{SnPc}, that of $C_{60}$ -- from \cite{C60}.

\section{Results and discussions}

\begin{figure}[h!]
\begin{center}
\resizebox*{10cm}{!}{\includegraphics{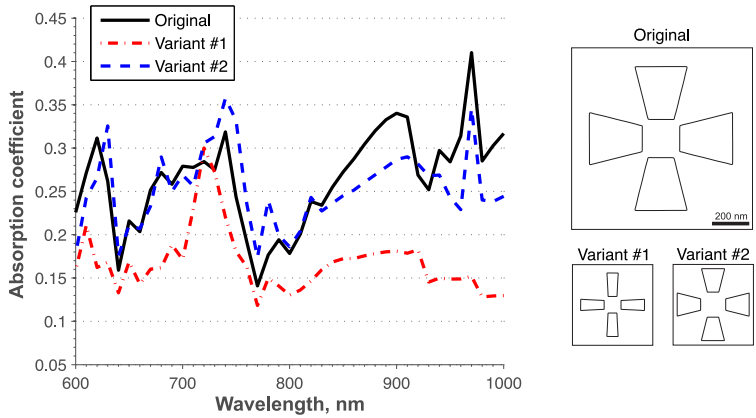}}
\caption{\label{fig2} Absorption coefficients as function of the wavelength for the initial design parameters of nanoantennas (black solid line), for narrowed nanoantennas (red dash dotted line), and for the case when nanostrips are shifted to the edges of the unit cell (blue dotted line). These geometries are depicted in scale on the right panel. The unit cell size and thickness of nanostrips are constant for all three cases and equal to 1000 nm and 50 nm, respectively. (The colour version of this figure is included in the online version of the journal.)}
\label{pic2}
\end{center}
\end{figure}

At first, numerical optimization of geometric parameters of nanoantennas was carried out to maximize the average photovoltaic absorption in the NIR range while maintaining the acceptable transparency of the whole solar cell for the visible light. Initial parameters of nanoantennas were taken the same as in \cite{nanoantennas}. Dramatic changes of horizontal sizes of nanoantennas are not required in this case, since the LTS was already scaled to the NIR in \cite{nanoantennas}. Therefore the optimization of nanoantennas reduces to finding the most advantageous arrangement of nanoantenna elements with respect to each other in the square unit cell of given area, to the optimization of lateral sizes of these elements and of their thickness. On the first stage we varied the maximal width of tapered nanostrips and the distance between them inspecting the distribution of the electric field in the layer of photoactive material and calculating the NIR absorption in the photovoltaic layer. Absorption coefficients of NIR solar light as functions of the wavelength for different horizontal parameters of nanoantennas and relative arrangement in the unit cell are shown in Fig. \ref{pic2}. Actually, the initial design with parameters from \cite{nanoantennas} turned out to be optimal. This result indicates the stability of domino modes with respect to the substrate properties. This robustness was assumed but not proved in \cite{nanoantennas}. Domino modes occupy the whole operation band 600-1000 nm whereas maxima and minima of the absorption coefficient correspond to constructive and destructive Fabry-Perot resonances, respectively.

\begin{figure}[h!]
\begin{center}
\includegraphics[width=8cm]{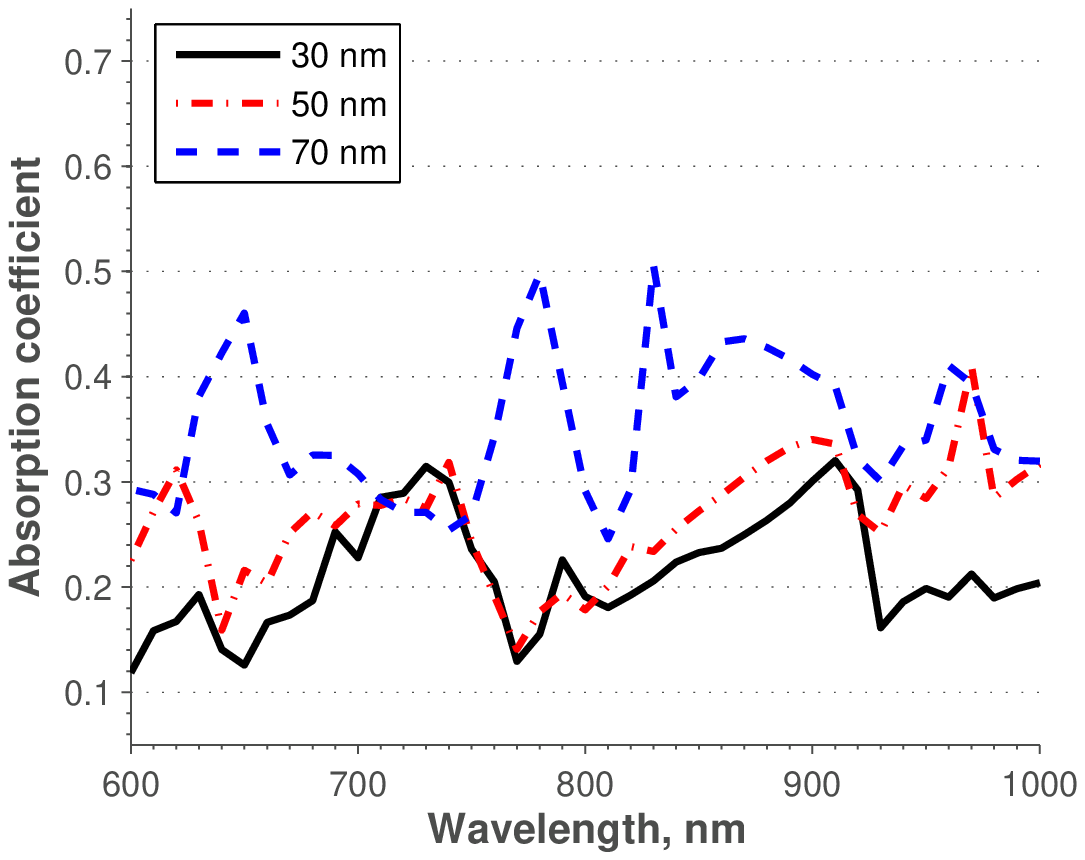}
\includegraphics[width=8cm]{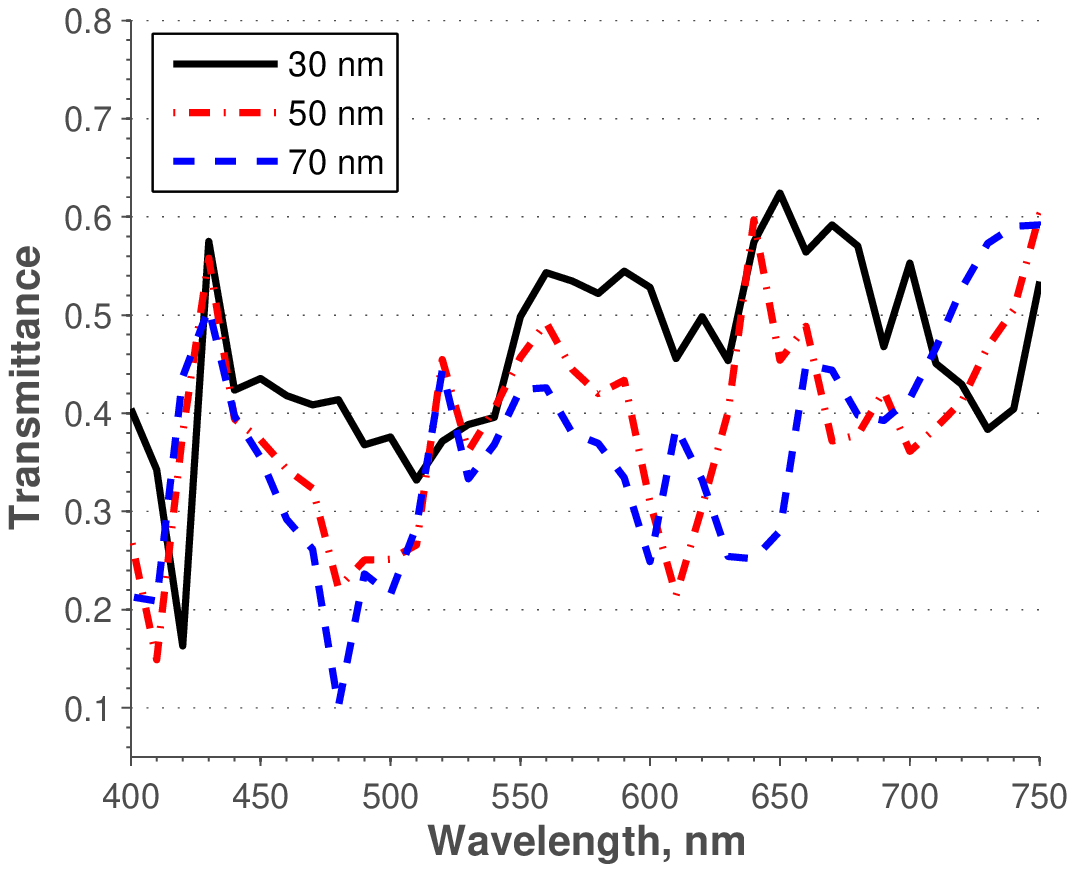}
\caption{\label{fig1} Absorption (left) and transmission (right) coefficients as functions of the wavelength for various thicknesses of metal nanoantennas: 30 nm, 50 nm (the same as in the case of the original design) and 70 nm. (The colour version of this figure is included in the online version of the journal.)}
\label{pic3}
\end{center}
\end{figure}

The second stage of optimization was related with maximization of AVT, for which the total area of metal elements per unit cell and their thickness are critical. The light trapping is better for thicker nanoantennas in which the domino modes are excited better. However, the thickness of metal elements has a negative impact on the AVT. The normalized infrared absorbance and visible transmittance as functions of the wavelength for various thicknesses of nanoantennas are shown in Fig. \ref{pic3}. The difference in the mean values of absorption in the NIR for 30 nm thick and 50 nm thick nanoantennas is not
very significant, whereas the difference in the AVT is strong. Considering the balance between the visible light transmittance and the NIR absorbance, the optimal thickness 30 nm was selected.

\begin{figure}[h!]
\begin{center}
\includegraphics[width=8cm]{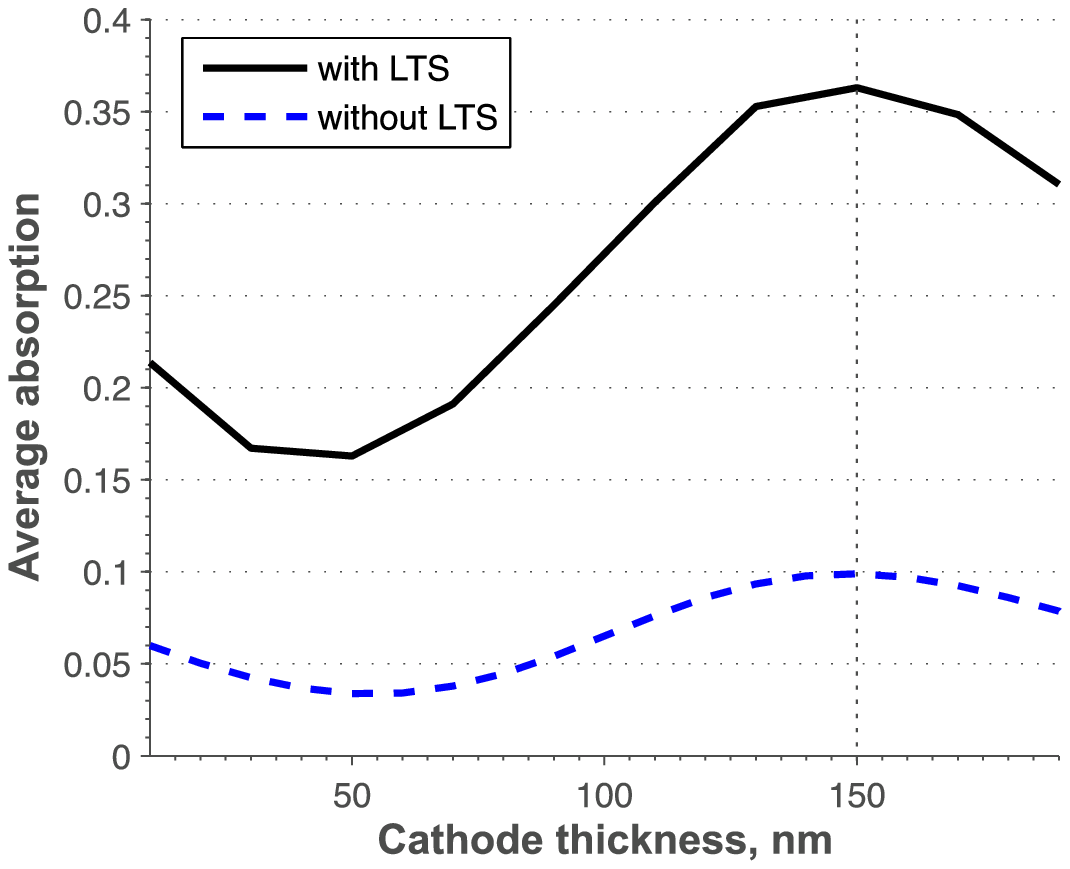}
\includegraphics[width=8cm]{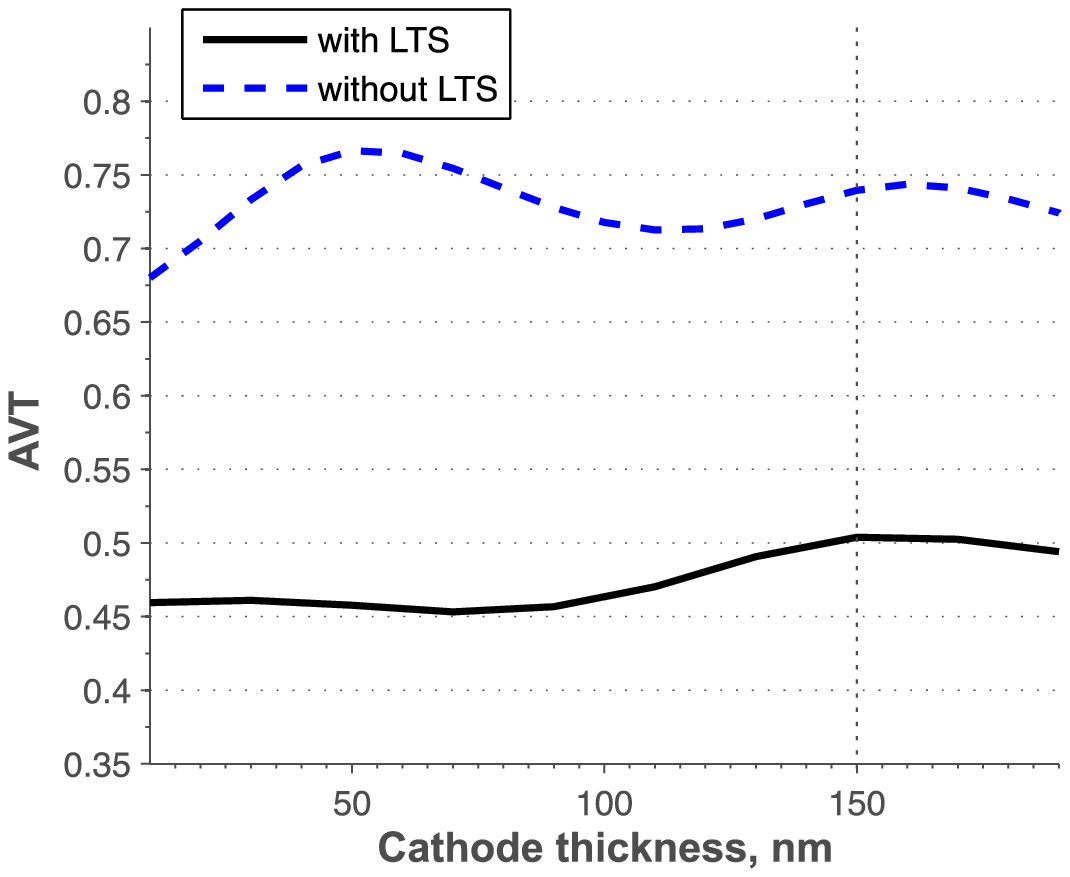}
\caption{\label{fig1} Average absorption (left) and average visible transmission (right) coefficients as functions of the wavelength for various cathode thicknesses. The optimal thickness of the transparent ITO cathode is 150 nm for both cases. (The color version of this figure is included in the online version of the journal.)}
\label{pic4}
\end{center}
\end{figure}

Next, we optimized the thickness of the ITO cathode keeping the constant thickness of the ITO anode equal to 150 nm and starting from 10 nm for the cathode thickness. These initial design parameters were taken from \cite{patent}. A proper choice of the cathode thickness enables to decrease the averaged reflectance in the operational band placing  constructive Fabry-Perot resonances of the overall structure inside it. These resonances also may decrease the transmittance maximizing the field in the active layer i.e. increasing the photovoltaic absorption. We calculated the field distributions and the photovoltaic absorption in the solar cell for different thicknesses of the ITO cathode. It was done for both pure solar cell and that with our LTS. Main results of these calculations are presented in Fig. \ref{pic4}.  The increase of the cathode thickness from initial 10 nm reduces the AVT rather slightly, whereas the positive impact of this increase to the NIR photovoltaic absorption is significant. The optimal thickness of the transparent ITO cathode was found 150 nm for a pure solar cell from the deal between the useful absorption and AVT. In this optimal case AVT is equal to 74\%.
In presence of our LTS, the optimal thickness of the ITO cathode turned out to be equal to 150 nm, too. This result indicates that nanoantennas practically do not contribute into wave interference processes. The AVT decreases from 74\% to 50\% due to the presence of nanoantennas with optimized parameters.
\begin{figure}[h!]
\begin{center}
\includegraphics[width=8cm]{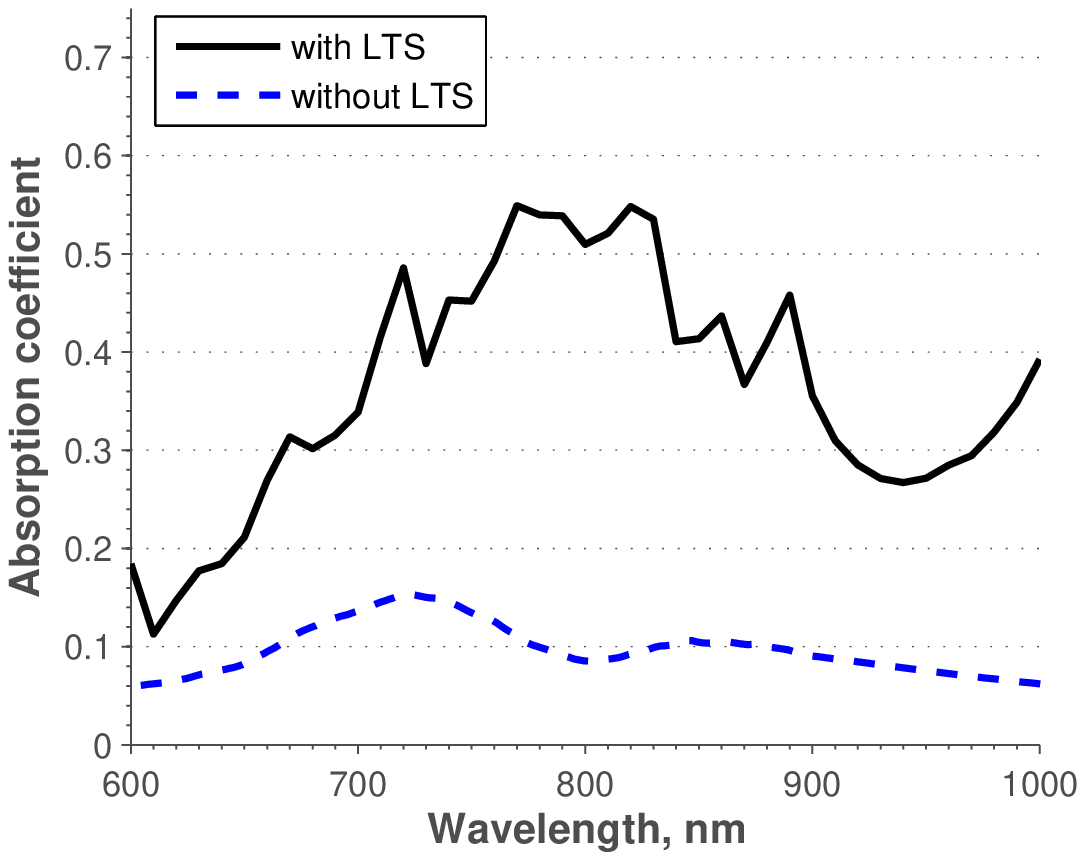}
\includegraphics[width=7cm]{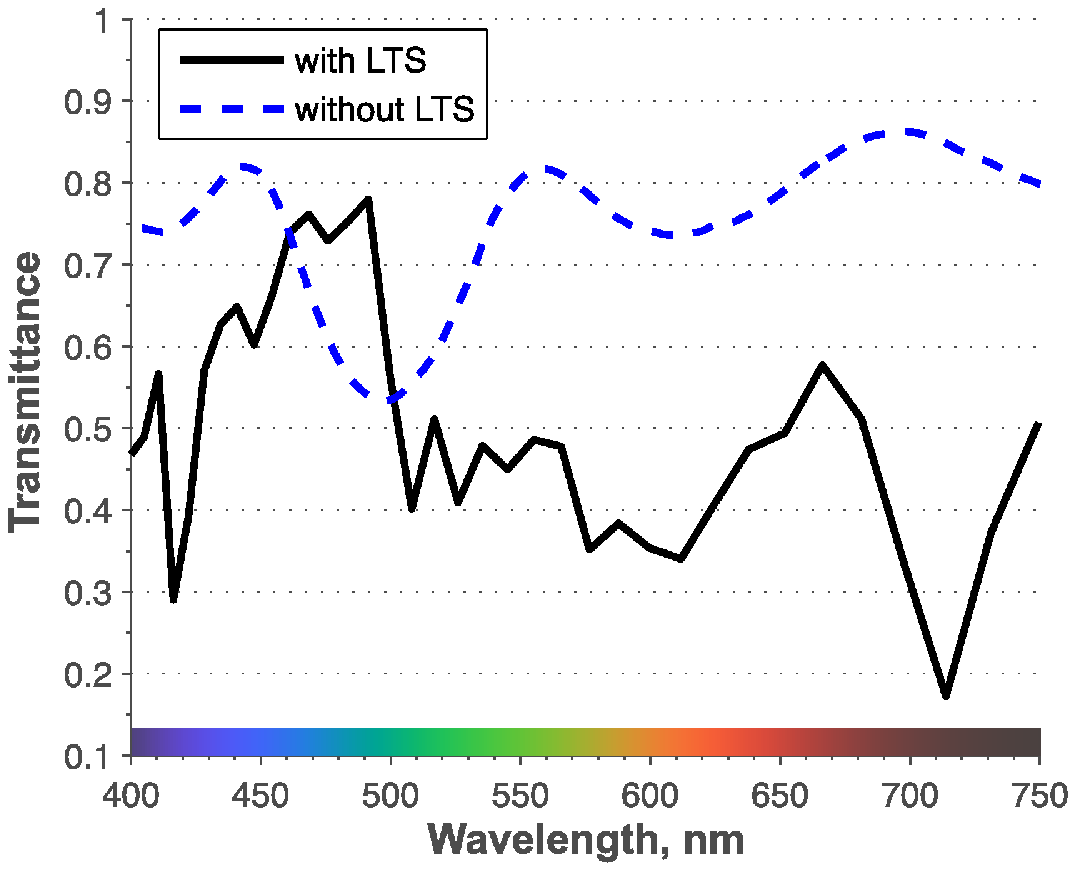}
\caption{\label{fig1} Absorption and transmission spectra in two cases: with our LTS and without LTS. (The colour version of this figure is included in the online version of the journal.)}
\label{pic5}
\end{center}
\end{figure}

Absorption and transmission spectra of the optimized structure are shown in Fig. \ref{pic5}. The total gain in the PV absorption granted by our LTS was calculated using the follow formula \cite{nanoantennas}:
\begin{equation}
G_A = \dfrac{J_{sc}^{LTS}}{J_{sc}^{\: w/o\: LTS}} = \dfrac{\int I_s(\lambda) R_s(\lambda) A_{LTS}(\lambda)\,d\lambda}{\int I_s(\lambda) R_s(\lambda) A_{\: w/o\: LTS}(\lambda)\,d\lambda}=3.6,
\label{abs}
\end{equation}
and the gain due to our LTS is more than triple. It is clear that this gain dominates over the rather modest reduction of the illumination. Notice, that in the present geometry the gain in the photovoltaic absorption is equal to the gain in the photocurrent \cite{Luke} and grants rather high electric power output.

\begin{figure}[h!]
\begin{center}
\includegraphics[width=8cm]{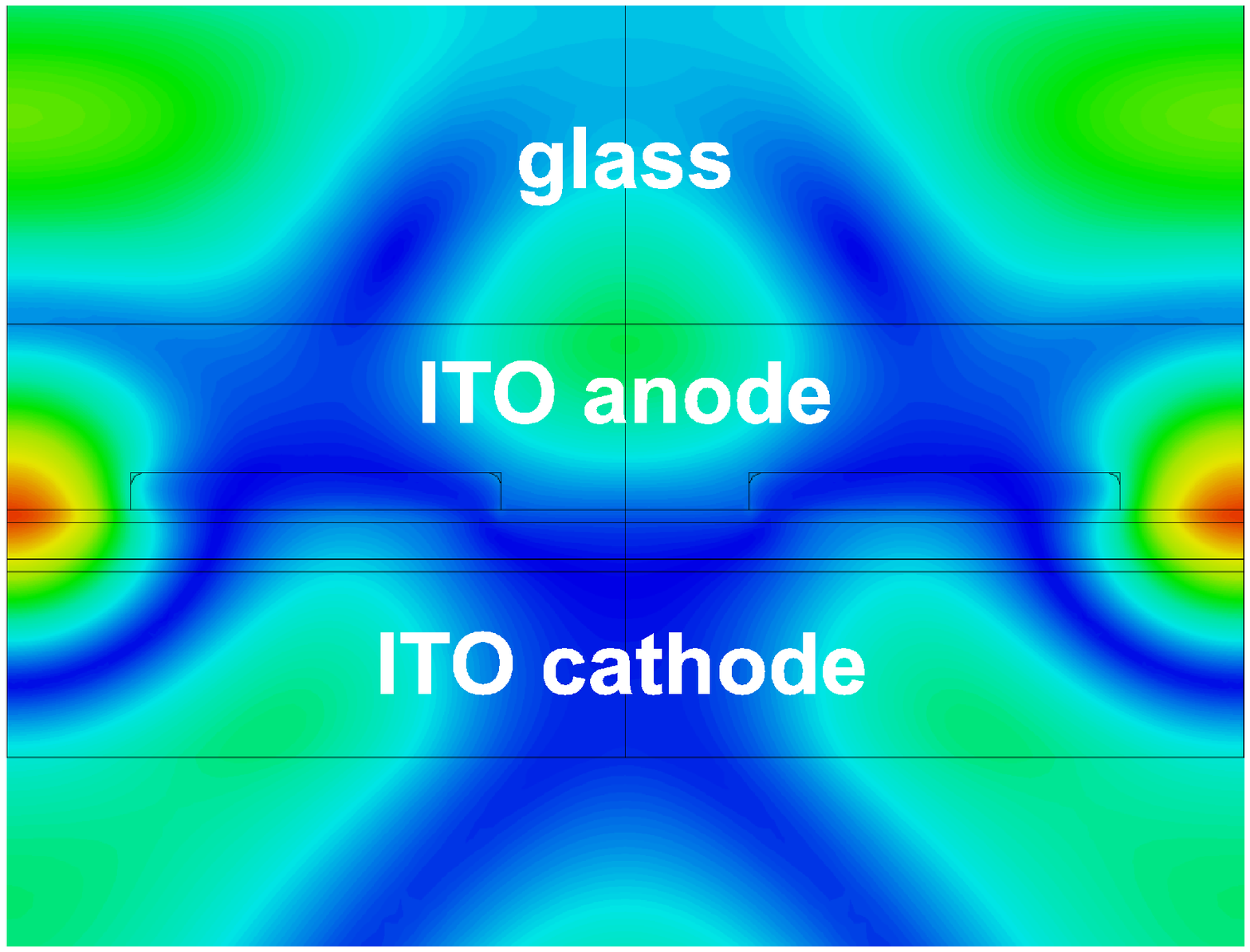}
\includegraphics[width=6.1cm]{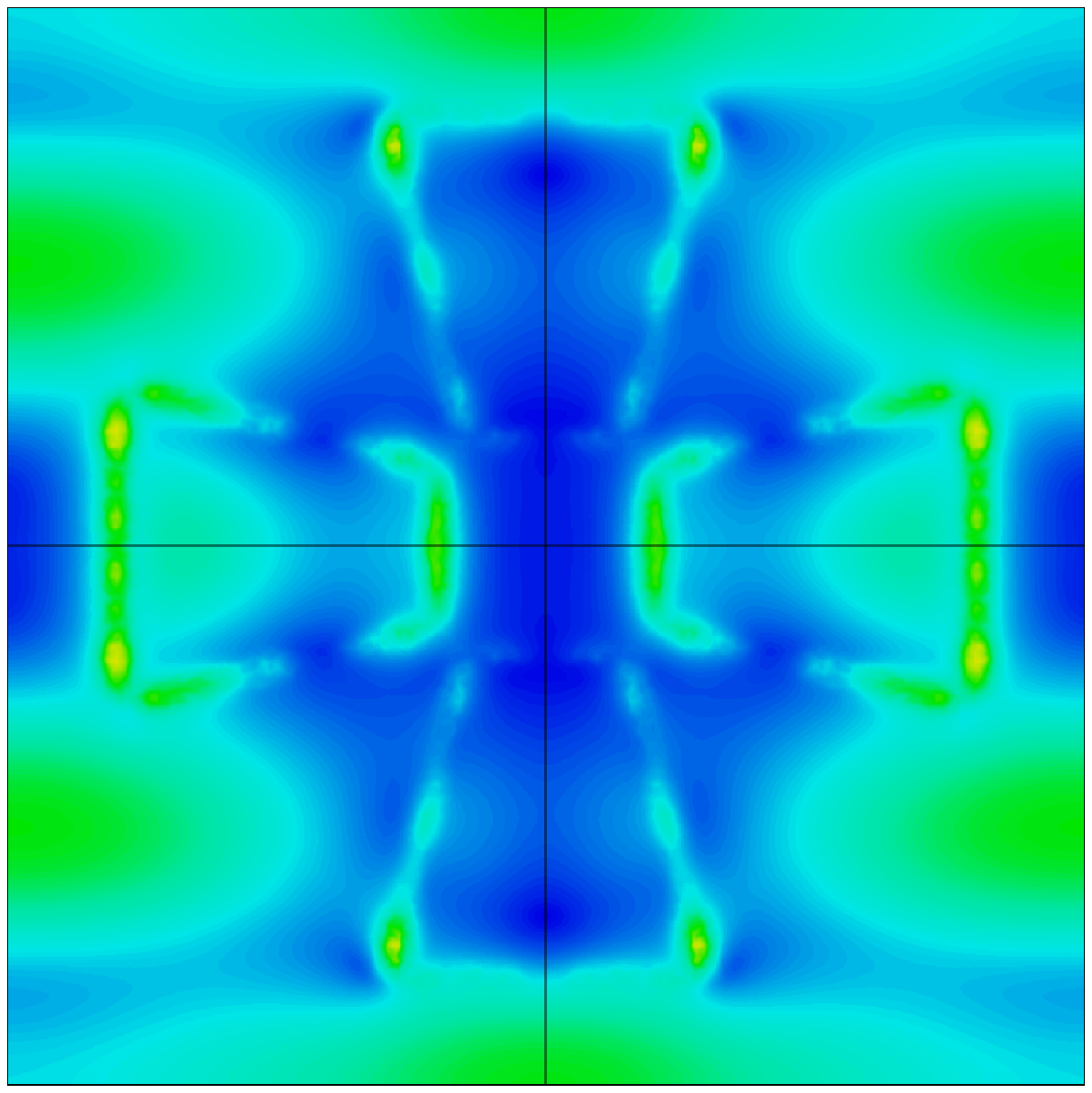}
\caption{\label{fig1} Electric field strength for $\lambda$ = 890 nm at the presence of the LTS in central vertical cross section (left) and horizontal plane at the interface between SnPc and $C_{60}$ (right). Hot spots produced by nanoantennas are partly located inside the active layers.  (The colour version of this figure is included in the online version of the journal.)}
\label{pic6}
\end{center}
\end{figure}

Electric field distributions in the vertical cross section of the structure and in the horizontal plane at the junction SnPc-C$_{60}$ for $\lambda=$875 nm are shown in Fig. \ref{pic6}. Similar distributions are typical for the whole range of domino modes 600-1000 nm.

\section{Conclusion}

In this paper we have shown that the light-trapping structure based on silver nanoantennas supporting so-called domino-modes can be used for the significant enhancement of photovoltaic absorption in organic solar cells with photovoltaic layers as thin as 40 nm. A practical design solution is suggested for an organic solar cell which operates in the NIR range and keeps rather transparent in the visible range. We studied the suggested structure and theoretically proved that domino modes are excited in the same range as they were predicted for same nanoantennas located on a substantial semiconductor substrate. We performed extended numerical simulations and optimized the structure achieving a significant (3.6 times) increase of photovoltaic absorption. The reduction of the transparency in the visible range from 74\% to 50\% due to the presence of nanoantennas is fully justified by more than triple gain expected for the photocurrent. The light-trapping structure is feasible \cite{nanoantennas}, and we envisage opportunities for the experimental demonstration of our claims. 

\section*{Acknowledgements}

This work has been partially supported by the grants of Russian Fund for Basic Research (number 13-08-01438 À, number 14-08-31730 mol\_a, number 14-02-31765 mol\_a), of the Government of the Russian Federation (Grant 074-U01), of the Ministry of Education and Science of Russian Federation (Project 11.G34.31.0020) and of the Dynasty Foundation.

\label{lastpage}

\end{document}